\documentclass[11pt, a4paper]{tboege-preprint}
\pdfoutput=1

\makeatletter
\@namedef{subjclassname@2020}{%
  \textup{2020} Mathematics Subject Classification}
\makeatother

\usepackage{multicol}
\usepackage{resizegather}
\usepackage{multibib}
\newcites{soft}{Mathematical software}

\renewcommand{\rho}{\varrho}
\newcommand{\Span}[1]{\langle#1\rangle}
\newcommand{\hrectangle}{{%
  \vcenter{\hbox{\scalebox{0.8}{%
    \ooalign{$\sqsubset\mkern11mu$\cr$\mkern11mu\sqsupset$\cr}%
  }}}%
}}

\title{No eleventh conditional Ingleton inequality}

\author{Tobias Boege}
\address{\mbox{Tobias Boege, Department of Mathematics, KTH Royal Institute of Technology, Sweden}}
\email{post@taboege.de}
\date{\today}


\begin{document}

\begin{abstract}
A rational probability distribution on four binary random variables
$X, Y, Z, U$ is constructed which satisfies the conditional independence
relations $\CI{X,Y}$, $\CI{X,Z|U}$, $\CI{Y,U|Z}$ and $\CI{Z,U|XY}$ but
whose entropy vector violates the Ingleton inequality.
This~settles a recent question of Studený (IEEE~Trans.\ Inf.\ Theory vol.~67,
no.~11) and shows that there are, up to symmetry, precisely ten inclusion-minimal
sets of conditional independence assumptions on four discrete random variables
which make the Ingleton inequality hold. The last case in the classification of
which of these inequalities are essentially conditional is also settled.
\end{abstract}

\maketitle

\section{Summary}

This note answers Open~Question~1 and one half of Open~Question~2 raised
by Milan Studený in his recent article \cite{StudenyIngleton} on
conditional Ingleton information inequalities on four discrete random
variables $X,Y,Z,U$. The first result is the following rational binary
distribution represented by its atomic probabilities $p_{ijk\ell} = {P(X=i, Y=j, Z=k, U=\ell)}$:
\begin{gather*}
  p_{0000} = \sfrac{20}{77}, \quad
  p_{0001} = 0, \quad
  p_{0010} = 0, \quad
  p_{0011} = 0, \\
  p_{0100} = \sfrac{20}{693}, \quad
  p_{0101} = \sfrac{4}{99}, \quad
  p_{0110} = \sfrac{10}{693}, \quad
  p_{0111} = \sfrac{2}{99}, \\
  p_{1000} = \sfrac{20}{693}, \quad
  p_{1001} = \sfrac{40}{99}, \quad
  p_{1010} = \sfrac{1}{693}, \quad
  p_{1011} = \sfrac{2}{99}, \\
  p_{1100} = 0, \quad
  p_{1101} = 0, \quad
  p_{1110} = 0, \quad
  p_{1111} = \sfrac{2}{11},
\end{gather*}
which satisfies solely the four conditional independence statements $\CI{X,Y}$,
$\CI{X,Z|U}$, $\CI{Y,U|Z}$ and $\CI{Z,U|XY}$ and on which the Ingleton
expression evaluates to a negative number close to~$-0.00757$.
This example settles simultaneously the last three open cases in the
classification of CI-type conditional Ingleton inequalities on four
discrete random variables and shows that all ten of them were already
described in~\cite{StudenyIngleton}.

With the knowledge of all conditional Ingleton inequalities, we also settle
the last remaining case in the classification of their essential conditionality.
The results are summarized in:

\begin{theorem*} \label{thm:Main}
On four discrete random variables $X, Y, Z, U$ there are precisely ten
inclusion-minimal conditional independence assumptions which make Ingleton's
inequality $\CIb{XY|ZU} \ge 0$ hold for entropy vectors (up to the symmetries
$X \leftrightarrow Y$ and $Z \leftrightarrow U$ in the Ingleton expression),
namely:
\begin{multicols}{2}
\vfill\noindent
\begin{gather*}
\label{eq:11cI} \tag{1.1} \CI{Z,U} \\
\label{eq:12cI} \tag{1.2} \CI{X,Z} \\
\label{eq:13cI} \tag{1.3} \CI{X,Z|Y} \\
\label{eq:14cI} \tag{1.4} \CI{X,Y|ZU} \\
\label{eq:15cI} \tag{1.5} \CI{X,Z|YU}
\end{gather*}
\vfill
\columnbreak
\vfill\noindent
\begin{align*}
\label{eq:21cI} \tag{2.1} \CI{X,Y}   &\wedge \CI{X,Y|Z} \\
\label{eq:22cI} \tag{2.2} \CI{X,Y|Z} &\wedge \CI{Y,U|Z} \\
\label{eq:23cI} \tag{2.3} \CI{X,Z|U} &\wedge \CI{X,U|Z} \\
\label{eq:24cI} \tag{2.4} \CI{X,Z|U} &\wedge \CI{Z,U|X} \\
\label{eq:25cI} \tag{2.5} \CI{X,Z|U} &\wedge \CI{Y,Z|U}
\end{align*}
\vfill
\end{multicols}
The conditional Ingleton inequalities given by \eqref{eq:11cI}--\eqref{eq:15cI}
are not essentially conditional but the ones given by \eqref{eq:21cI}--\eqref{eq:25cI}
are essentially conditional.
\end{theorem*}

These results are derived computationally. \Cref{sec:Ingleton} gives an
introduction to the topic of conditional Ingleton inequalities and recalls
the previous results leading to the question answered here. For basics on
polymatroids and their role in conditional independence and information
theory we refer to the excellent exposition in \cite{StudenyIngleton}.
The~computational methodologies used to find the above distribution and
to prove essential conditionality of inequality \eqref{eq:25cI} are
explained in \Cref{sec:Construction,sec:Essential}, respectively.
\Cref{sec:Remarks} collects further remarks and observations.
The~source code in \TT{Macaulay2} \citesoft{M2} and \TT{Mathematica}
\citesoft{Mathematica} behind various steps in the computations and
auxiliary data produced using \TT{4ti2} \citesoft{4ti2} and \TT{normaliz}
\citesoft{Normaliz} are available at
\begin{center}
  \url{https://mathrepo.mis.mpg.de/ConditionalIngleton/}.
\end{center}

\section{On conditional Ingleton inequalities}
\label{sec:Ingleton}

\subsection{Ingleton inequality and entropy region}
\label{sec:Entropy}

Suppose that $X,Y,Z,U$ are subspaces in a finite-dimensional (left
or right) vector space over a field (or division ring). For this data,
the \emph{Ingleton inequality} asserts that
\begin{equation}
  \label{eq:Ingleton}
  \tag{$\CIbOp$}
  \begin{aligned}
  0 &\le \dim\Span{X,Z} + \dim\Span{X,U} + \dim\Span{Y,Z} + \dim\Span{Y,U} + \dim\Span{Z,U} - {} \\
    &\hphantom{{}\!\le{}} \hphantom{-} \dim\Span{X,Y} - \dim\Span{Z} - \dim\Span{U} - \dim\Span{X,Z,U} - \dim\Span{Y,Z,U},
  \end{aligned}
\end{equation}
where $\dim\Span{\blank}$ is the dimension of the subspace spanned
by its arguments. This rank function $h: 2^N \to \BB R$,
$I \mapsto \dim\Span{I}$ on subsets of $N = \Set{X,Y,Z,U}$ is a
\emph{polymatroid}, i.e., it is normalized: $h(\emptyset) = 0$,
non-decreasing: $h(I) \le h(J)$ for $I \subseteq J$, and submodular:
$h(I) + h(J) \ge h(I \cup J) + h(I \cap J)$ for all $I, J \subseteq N$.
The set of polymatroids over an $n$-element set forms a rational
polyhedral cone in $\BB R^{2^n}$ denoted by~$\BO H_n$.
The \emph{Ingleton expression} $\CIb{XY|ZU}$ is the linear functional
in $h$ which appears on the right-hand side of \eqref{eq:Ingleton}.
Hence, non-negativity of the inner product $\CIb{XY|ZU} \cdot h$ is
a necessary condition for a polymatroid~$h$ to be linearly representable
over some division ring.
The necessity was found by Ingleton \cite{Ingleton} through an analysis
of the Vámos matroid, the prototypical example of a non-linear matroid.

Now let $X,Y,Z,U$ denote jointly distributed random variables which
take only finitely many states. These random variables are referred
to as \emph{discrete} with finiteness being implicit. If $X$ attains
$q$ states, without loss of generality from the set $[q] \defas
\Set{1, \dots, q}$, with positive probabilities $p(X=i)$, then its
\emph{Shannon entropy} is the expression
\[
  H(X) \defas \BB E_X[\log\sfrac1{p}] = \sum_{i=1}^q p(X=i) \log \sfrac1{p(X=i)}.
\]
The \emph{entropy vector} of jointly distributed discrete random variables
$X_1, \dots, X_n$ assigns to each subset $I \subseteq [n]$ the entropy of
the vector-valued discrete random variable $X_I \defas (X_i : i \in I)$.
We~denote the \emph{entropy region}, the set of all points in $\BB R^{2^n}$
which occur as entropy vectors of $n$ discrete random variables, by~$\BO H_n^*$.
The choice of basis for the logarithm changes the scale of all entropy vectors
and does not change any of the considerations in this paper.

Fujishige \cite{Fujishige} observed that the non-negativity of Shannon's
information measures implies that entropy vectors are polymatroids and
thus entropy vectors are sometimes called \emph{entropic polymatroids}.
A result of Matúš \cite[Lemma~10]{MatusMinors} implies that every integer
polymatroid which is linearly representable by a subspace arrangement over
a field is a scalar multiple of an entropic one.
Hence, it makes sense to reinterpret Ingleton's functional $\CIb{XY|ZU}$
by replacing $\dim\Span{\blank}$ with $H(\blank)$. But whereas the
inequality $\CIbOp \ge 0$ is valid for linear polymatroids, it fails
for the more general entropic ones. This~paper is concerned with
special types of assumptions on entropy vectors which guarantee that
the Ingleton inequality~holds.

\subsection{Discrete representability of CI~structures}
\label{sec:CI}

Even though the Ingleton inequality does not hold universally for entropy
vectors, it was a key tool in the characterization of \emph{conditional
independence (CI) structures} which are representable by four discrete
random variables. This classification was achieved in the series of
papers \cite{MatusI,MatusII,MatusIII} by Matúš and Studený and we use
this section to outline the role of the Ingleton inequality in this work.

Let $I,J,K \subseteq [n]$. The common shorthand notation $IJ \defas I \cup J$
applies to these subsets. For a polymatroid~$h$ and $I,J,K \subseteq [n]$,
we employ the \emph{difference expression}
\[
  \CId{I,J|K}\cdot h \defas h(IK) + h(JK) - h(IJK) - h(K),
\]
that is, $\CId{I,J|K}$ is a linear functional on $\BB R^{2^n}$.
The non-negativity of this functional on $\BO H_n$ is guaranteed by the
submodular inequalities. Its vanishing makes $IK$ and $JK$ a \emph{modular pair}.
If $h$ is the entropy vector of random variables $(X_i : i \in [n])$,
then $\CId{I,J|K}$ is known as the \emph{conditional mutual information}
of subvectors $X_I$ and $X_J$ given $X_K$ and its vanishing is equivalent
to the conditional independence~$\CI{X_I,X_J|X_K}$.
Recall from \cite[Section~II.D]{StudenyIngleton} that the study of
conditional independence (excluding functional dependence) can be
reduced to the \emph{elementary CI~statements}, i.e., the equalities
$\CId{i,j|K} = 0$ where $i$ and $j$ are distinct singletons and $K$
is a subset of $N$ not containing $i$ or~$j$.
These functionals define facets of $\BO H_n$ and even supporting
hyperplanes of $\BO H_n^*$ with non-empty intersection.
A set $\CC L$ of elementary CI~statements on $n$ random variables, also
called a \emph{CI~structure}, is \emph{representable} if and only if there
exists $h \in \BO H_n^*$ such that $\CId{i,j|K}\cdot h = 0 \;\Leftrightarrow\;
\CI{i,j|K} \in \CC L$. The CI~structure defined by any polymatroid $h$
in this way is denoted by $\CIS{h}$.

Let $\BO H_4^\CIbOp$ denote the subcone of $\BO H_4$ (whose ground set
elements are labeled $X,Y,Z,U$) which consists of polymatroids satisfying
the Ingleton inequalities $\CIb{IJ|KL} \ge 0$ for every possible permutation
$I,J,K,L$ of $X,Y,Z,U$. There are $\binom42 = 6$ unique such inequalities
because the Ingleton expression is invariant under exchanging $I \leftrightarrow J$
and $K \leftrightarrow L$.
One~key insight of \cite{MatusI} is that the extreme rays of $\BO H_4^\CIbOp$
are a subset of those of $\BO H_4$ and that they are all probabilistically representable.
This implies that every CI~structure $\CIS{h}$, for $h \in \BO H_4^\CIbOp$,
is representable; this condition is of polyhedral nature and can easily be
checked using linear programming.
Miraculously, even in the non-Ingleton regime, the Ingleton inequality is
the main obstruction to entropicness: namely, in \cite{MatusIII} sets of
CI~statements $\CC L$ are described such that whenever a polymatroid $h$ is
entropic and satisfies $\CC L \subseteq \CIS{h}$, then $\CIb{XY|ZU} \cdot h
\ge 0$ holds. This is a conditional information inequality in the sense of
\cite{CondInfo}, formally written as $\CC L \Rightarrow \CIb{XY|ZU} \ge 0$
and called a \emph{conditional Ingleton inequality}. It is important to
emphasize that a conditional information inequality is not required to hold
for general polymatroids (in which case it would be a consequence of the
polyhedral geometry of~$\BO H_4$ and not very informative) but only for
entropic polymatroids. An inequality such as $\CC L \Rightarrow \CIb{XY|ZU} \ge 0$
allows one to conclude that a CI~structure containing $\CC L$ cannot be
representable if the cone of its realizing polymatroids does not intersect
the cone given by $\CIb{XY|ZU} \ge 0$; which is again a polyhedral condition
that can be computed easily.

\begin{convention*}
The definition of conditional information inequality in \cite{CondInfo}
allows arbitrary linear assumptions $p_1 \cdot h \ge 0 \wedge \cdots \wedge
p_s \cdot h \ge 0$ to imply a linear conclusion $q \cdot h \ge 0$.
Conditional independence assumptions are a special case of this using
$\CIdOp$ functionals. In this work, ``conditional information inequality''
will always refer to the special case of CI-type inequality.
\end{convention*}

While the precise shape of $\BO H_4^*$ or even its closure $\ol{\BO H_4^*}$
in the euclidean topology (which is known to be a convex cone \cite{NonShannon})
remains unknown to date (cf.\ \cite{MatusInfinite} and \cite{GMM} for a
challenging open problem), conditional information
inequalities help to delimit it in ways that go beyond linear inequalities
and hence make it possible to describe differences between the entropy region
and its closure. This becomes significant, for example, when
information-theoretic optimization problems such as channel capacity
computations are solved not in terms of their original parameters and
non-linear objective functions but in terms of linear programs over
the entropy region; this is done in Shannon's original paper \cite[Theorem~10]{Shannon}
and has since then become a standard technique.
In this case, the optimum is attained on the boundary of $\ol{\BO H_n^*}$.
Even if it can be located, it is not clear whether the optimizer is
entropic and hence corresponds to a real probability distribution or
if it can only be approximated arbitrarily well by distributions.

The knowledge of which CI~structures are representable can be viewed
as combinatorial information about the intricate boundary structure
of $\BO H_4^*$. Namely, given a set of CI~assumptions~$\CC L$ which
define a subspace $U = \Set{ h \in \BB R^{16}: \CId{i,j|K} \cdot h = 0
\; \text{for all $\CI{i,j|K} \in \CC L$} }$, the question is which
other inequalities $\CIdOp \ge 0$ are tight at every point in $\BO H_n^* \cap U$?
Calling the set of implied statements $\CC M$, this proves a
\emph{conditional independence inference rule} $\CC L \Rightarrow \CC M$
for representable CI~structures. Unlike the geometric shape of $\BO H_4^*$,
this combinatorial, CI-theoretic information about its boundary is
completely available due to the series of papers by Matúš and Studený.
Studený's recent paper \cite{StudenyIngleton} revisits this series
and shows that all inference properties for four discrete random
variables can be deduced from conditional Ingleton inequalities
in addition to the common Shannon information inequalities.
Each of the ten conditional Ingleton inequalities presented in
\cite{StudenyIngleton} is necessary to obtain all the CI~inference
rules. In this paper we prove that there are no further, in the
CI-theoretic sense ``extraneous'', conditional Ingleton inequalities.

\subsection{Masks and conditional Ingleton inequalities}
\label{sec:CondIng}

One way to obtain conditional Ingleton inequalities is to rewrite the
functional $\CIb{XY|ZU}$ as a linear combination of difference expressions
$\CId{i,j|K}$ in the dual space $(\BB R^{16})^*$. Some of these \emph{masks}
of the Ingleton expression were found in \cite{MatusI} and are also
discussed in \cite[Section~II.G]{StudenyIngleton}:
\begin{align}
  \CIb{XY|ZU}
  &= \CId{Z,U|X} + \CId{Z,U|Y} + \CId{X,Y}    - \CId{Z,U}     \tag{M.1} \label{eq:M:1} \\
  &= \CId{Z,U|Y} + \CId{X,Z|U} + \CId{X,Y}    - \CId{X,Z}     \tag{M.2} \label{eq:M:2} \\
  &= \CId{X,Y|Z} + \CId{X,Z|U} + \CId{Z,U|Y}  - \CId{X,Z|Y}   \tag{M.3} \label{eq:M:3} \\
  &= \CId{X,Y|Z} + \CId{X,Y|U} + \CId{Z,U|XY} - \CId{X,Y|ZU}  \tag{M.4} \label{eq:M:4} \\
  &= \CId{X,Y|Z} + \CId{X,Z|U} + \CId{Z,U|XY} - \CId{X,Z|YU}. \tag{M.5} \label{eq:M:5}
\end{align}
These masks prove \eqref{eq:11cI}--\eqref{eq:15cI}; indeed mask~\eqref{eq:M:1},
for example, implies \eqref{eq:11cI}: $\CI{Z,U} \Rightarrow \CIb{XY|ZU} \ge 0$
due to the non-negativity of all difference expressions. Under the symmetries
$X \leftrightarrow Y$ and $Z \leftrightarrow U$ which fix $\CIb{XY|ZU}$, these
five masks generate fourteen distinct conditional Ingleton inequalities,
displayed below in groups by symmetry class:
\begin{gather*}
  \Big(\CI{Z,U}\Big), \quad
  \Big(\CI{X,Z}, \;\;
  \CI{Y,Z}, \;\;
  \CI{X,U}, \;\;
  \CI{Y,U}\Big), \\
  \Big(\CI{X,Z|Y}, \;\;
  \CI{Y,Z|X}, \;\;
  \CI{X,U|Y}, \;\;
  \CI{Y,U|X}\Big), \quad
  \Big(\CI{X,Y|ZU} \Big), \\
  \Big(\CI{X,Z|YU}, \;\;
  \CI{Y,Z|XU}, \;\;
  \CI{X,U|YZ}, \;\;
  \CI{Y,U|XZ}\Big).
\end{gather*}
In \cite[Section~IV]{StudenyIngleton} five further conditional Ingleton
inequalities are proved which require \emph{two} CI~assumptions. They
expand to fourteen conditional inequalities under symmetry as well.
Studený has ruled out five other sets of CI~assumptions by counterexamples
and reduced the possibilities for an eleventh conditional Ingleton inequality
to three CI~structures, namely the sets strictly above $\CC L_0 = \CI{X,Z|U}
\wedge \CI{Y,U|Z}$ and below $\CC L = \CI{X,Z|U} \wedge \CI{Y,U|Z} \wedge
\CI{X,Y} \wedge \CI{Z,U|XY}$.

The verification of this claim by hand is tedious. The process can be
delegated to a \TT{SAT} solver such as \TT{CaDiCaL}~\citesoft{CaDiCaL}
as follows. There are 24 elementary CI~statements $\CI{i,j|K}$ on four
random variables; introduce one boolean variable for each of them.
If a CI~structure implies the Ingleton inequality, then so does every superset.
If a counterexample exists for a set of CI~assumptions, then every subset
is ruled out by the same counterexample. Using the ten known conditional
Ingleton inequalities, Studený's five counterexamples and the conjectured
minimal and maximal unsolved cases $\CC L_0$ and $\CC L$ --- and all their
symmetric variants ---, a boolean formula can be constructed whose satisfying
assignments are all CI~structures which are not covered and are potential
assumptions for an eleventh conditional Ingleton inequality. The solver
quickly decides that the formula is unsatisfiable and hence proves that
all unsolved cases are between $\CC L_0$ and $\CC L$. More details and
source code for this computation are available on our MathRepo~page.

The objective of the next section is to construct a probability distribution
satisfying $\CC L$ and violating the Ingleton inequality.
Known examples of this kind are usually hand-crafted, rational distributions
with small denominators derived by careful exploitation of zero patterns and
symmetries; cf.~\cite{CondInfo,StudenyIngleton}.
We present a different, computer-assisted and heuristic methodology to find
counterexamples in information theory rooted in algebra and relying on
symbolic computations as well as numerical non-linear optimization.

\section{Construction of the distribution}
\label{sec:Construction}

\subsection{Circuits, masks and scores}
\label{sec:Masks}

The $24$ difference expressions $\CId{i,j|K}$ and the Ingleton expression
$\CIb{XY|ZU}$ are elements in the dual space $(\BB R^{16})^*$. Choosing
the standard basis there, they can be identified with vectors which make
up the columns of a $16 \times 25$ matrix. The circuits of this matrix,
i.e., the non-zero integer vectors in its kernel with inclusion-minimal
support and coprime non-zero entries, can be computed using the software
\TT{4ti2} \citesoft{4ti2}; cf.~\cite[Chapter~4]{SturmfelsConvex}.
There are $10\,481$ such circuits and among them $6\,814$ which give
a non-zero coefficient to $\CIb{XY|ZU}$. These circuits are the shortest
possible ways of writing $\CIbOp$ as a linear combination of~$\CIdOp$.
The 14 shortest circuits require only four $\CIdOp$ terms one of which
with a negative coefficient; they are precisely the 14 symmetric
images of \eqref{eq:M:1}--\eqref{eq:M:5}. All masks are available on
our~website.

Based on the circuits, we obtain short masks which are closely
related to the two subcases $\CC L_1 = \CI{Z,U|XY} \wedge \CC L_0$
and $\CC L_2 = \CI{X,Y} \wedge \CC L_0$ of the model $\CC L =
\CC L_1 \wedge \CC L_2$. All three cases remained open in Studený's
analysis, but $\CC L_0$ was settled in \cite[Example~5]{StudenyIngleton}.
The mask
\begin{align}
  \label{eq:Mask:1} \tag{$\dagger_1$}
  \begin{split}
    \CIb{XY|ZU} &= \CId{X,Y|ZU} + \CId{X,Z|U} - \CId{X,Z|YU} + \CId{Y,U|Z} - {} \\
                &\hphantom{{}={}} \CId{Y,U|XZ} + \CId{Z,U|XY},
  \end{split}
\end{align}
can be confirmed by plugging in the definitions of $\CIdOp$ and $\CIbOp$.
It was selected to simplify as much as possible under the CI~assumptions
$\CC L_1$ which would otherwise contribute positive quantities to the
Ingleton expression. Given that $\CC L_1$ holds, the mask~\eqref{eq:Mask:1}
yields
\begin{equation}
  \label{eq:Score:1} \tag{$\ddagger_1$}
  \begin{split}
    -\CIb{XY|ZU} &= \CId{X,Z|YU} + \CId{Y,U|XZ} - \CId{X,Y|ZU} \\
                 &= H(Y|XZ) + H(X|YU) - H(XY|ZU) \asdef \rho_1(X,Y,Z,U).
  \end{split}
\end{equation}
Analogously one proves
\begin{align}
  \label{eq:Mask:2} \tag{$\dagger_2$}
  \CIb{XY|ZU} &= \CId{X,Y} - \CId{X,Z} + \CId{X,Z|U} - \CId{Y,U} + \CId{Y,U|Z} + \CId{Z,U},
\end{align}
which under $\CC L_2$ yields
\begin{equation}
  \label{eq:Score:2} \tag{$\ddagger_2$}
  \begin{split}
    -\CIb{XY|ZU} &= \CId{X,Z} + \CId{Y,U} - \CId{Z,U} \\
                 &= H(ZU) - H(Z|X) - H(U|Y) \asdef \rho_2(X,Y,Z,U).
  \end{split}
\end{equation}
The functions $\rho_1$ and $\rho_2$ are referred to as the
\emph{non-Ingleton scores} on $\CC L_1$ and $\CC L_2$, respectively.
On~the distributions satisfying the respective CI~statements, they equal
the value of $-\CIb{XY|ZU}$ but they involve fewer terms and are thus
easier to evaluate and to differentiate. Both scores coincide on
the intersection $\CC L$ of the models $\CC L_1$ and $\CC L_2$.

We continue with a geometric analysis of the space of binary distributions
in the model $\CC L_1$ and extend these findings to derive a binary
distribution for $\CC L$ with positive non-Ingleton score.

\subsection{Parametrization of $\CC L_1$}

A joint distribution of four binary random variables is given by a
$2 \times 2 \times 2 \times 2$ tensor with real, non-negative entries
$p_{ijk\ell}$ which sum to one. With all four indices ranging in
$\Set{0,1}$, these represent the atomic probabilities of the sixteen
joint events. The~CI~statements of $\CC L_1$ prescribe quadratic
equations on these probabilities:
\begin{align*}
  \CI{Z,U|XY} &\;\Leftrightarrow\; \begin{cases}
    p_{0000} \cdot p_{0011} = p_{0001} \cdot p_{0010}, \\
    p_{0100} \cdot p_{0111} = p_{0101} \cdot p_{0110}, \\
    p_{1000} \cdot p_{1011} = p_{1001} \cdot p_{1010}, \\
    p_{1100} \cdot p_{1111} = p_{1101} \cdot p_{1110},
  \end{cases} \\[0.5em]
  \CI{X,Z|U} &\;\Leftrightarrow\; \begin{cases}
    (p_{0000} + p_{0100}) \cdot (p_{1010} + p_{1110}) = (p_{0010} + p_{0110}) \cdot (p_{1000} + p_{1100}), \\
    (p_{0001} + p_{0101}) \cdot (p_{1011} + p_{1111}) = (p_{0011} + p_{0111}) \cdot (p_{1001} + p_{1101}),
  \end{cases} \\[0.5em]
  \CI{Y,U|Z} &\;\Leftrightarrow\; \begin{cases}
    (p_{0000} + p_{1000}) \cdot (p_{0101} + p_{1101}) = (p_{0001} + p_{1001}) \cdot (p_{0100} + p_{1100}), \\
    (p_{0010} + p_{1010}) \cdot (p_{0111} + p_{1111}) = (p_{0011} + p_{1011}) \cdot (p_{0110} + p_{1110}).
  \end{cases}
\end{align*}

\pagebreak

These equations are studied in algebraic statistics; see \cite[Proposition~4.1.6]{Sullivant}
for their derivation. It is in general difficult to derive a rational
parametrization of a given CI~model. To~simplify this task, we impose
the support pattern which already appears in \cite[Example~5]{StudenyIngleton}:
suppose that $p_{0001} = p_{0010} = p_{0011} = p_{1100} = p_{1101} = p_{1110} = 0$
and all other variables are positive. From now on, we regard only this
linear slice of the CI~models for $\CC L_1$, $\CC L_2$ and $\CC L$.

Under these additional constraints, the above eight equations together with
the condition that all probabilities sum to one can be resolved to yield the
rational parametrization
\[
  \label{eq:Param} \tag{$*$}
  \begin{gathered}
    p_{0100} = \frac{p_{0101} p_{0110}}{p_{0111}}, \quad
    p_{1000} = \frac{p_{1001} p_{1010}}{p_{1011}}, \\
    p_{0111} = \frac{p_{0101}}{p_{1001} (p_{1011} + p_{1111})}, \quad
    p_{0000} = \frac{p_{0110} p_{1001} p_{1111}}{p_{1011} (p_{1011} + p_{1111})}, \\
    p_{1010} = \frac{p_{0110} p_{1011}}{p_{1011} + p_{1111}}, \quad
    p_{0101} = \frac{p_{1001} p_{1011}}{p_{1011} + p_{1111}}, \\
    p_{1001} = \frac{p_{1011}^2 (1 - 2 p_{0110} - 2 p_{1011}) +
      p_{1011} p_{1111} (1 - p_{0110} - 3 p_{1011} - p_{1111})}%
      {(p_{0110} + p_{1011}) (2 p_{1011} + p_{1111})}.
  \end{gathered}
\]
With six zero conditions and seven equations (two of the CI~equations trivialize
under the zero constraints), this leaves the three parameters $p_{0110}$, $p_{1011}$
and $p_{1111}$. The positivity conditions on the ten non-zero probabilities turn
into non-linear inequalities and these are the only remaining constraints on the
parameters. Thus, this defines a three-dimensional basic semialgebraic set~$\CC T_1$.

\begin{figure}
\includegraphics[width=0.8\textwidth]{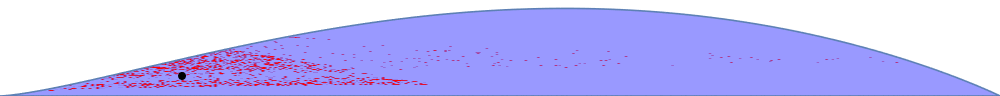}
\caption{The model $\CC L$ in its $(p_{1111}, p_{1011})$-parameter space
$\CC T$. Points with a positive non-Ingleton score $\rho_2$ are colored
in red. The rational non-Ingleton distribution with $p_{1011} = \sfrac{2}{99}$
and $p_{1111} = \sfrac{2}{11}$ is marked with a black dot.}
\label{fig:Score}
\end{figure}

\subsection{Numerical optimization and a rational point}

The Ingleton inequality is not an algebraic function of the parameters
but a transcendental one. Hence, algebraic techniques like Gröbner bases
or cylindrical algebraic decomposition cannot be directly applied to
decide if there exist parameters on which $\CIb{XY|ZU}$ is negative.
This~question can be reformulated as whether a system of integer polynomial
equations and inequalities in variables and exponentials of variables has
a real solution. Thus, it is a question in the existential theory of the
real numbers with exponentiation. The decidability of this theory is
an open problem known as Tarski's Exponential Function Problem and hence
no general symbolic algorithms are available today to solve it;
see~\cite{RealExponential} for a starting point on this topic.

Instead of symbolic techniques, we employ optimization. \TT{Mathematica}'s
\TT{FindMaximum} function, when started on the values $(\sfrac{1}{16},
\sfrac{1}{16}, \sfrac{1}{16})$, numerically finds a local maximum of $\rho_1$
on~$\CC T_1$ with value $0.0198$ at the parameters $p_{0110} = 0.36179$,
$p_{1011} = 0.01463$ and $p_{1111} = 0.27455$. By~continuity, $\rho_1$
remains positive in a small neighborhood of this point. Searching for a
\emph{local minimum} of $\rho_1$ in the range
\[
  \label{eq:Bounds} \tag{$\protect\hrectangle$}
   \sfrac16     \le p_{0110} \le \sfrac36,     \quad
   \sfrac1{160} \le p_{1011} \le \sfrac3{160}, \quad
   \sfrac18     \le p_{1111} \le \sfrac38
\]
yields a positive value, indicating that this region is likely to contain
many points violating the Ingleton inequality. Based on this heuristic,
we want to find a distribution in this range which satisfies the system $\CC T$
consisting of the inequalities of $\CC T_1$ and the additional CI~equation
for $\CI{X,Y}$ which rewrites under the parametrization~\eqref{eq:Param} to
\begin{gather*}
  p_{1011}^{2} (p_{1011} + p_{1111})^{3} + p_{0110}^{2} p_{1111} (2 p_{1011}^{3} + p_{1111}^{4} + p_{1011} p_{1111}^{2} (1 + 4 p_{1111}) + p_{1011}^{2} p_{1111} (3 + 4 p_{1111})) + {} \\
  p_{0110} (p_{1011}^{4} + 5 p_{1011} p_{1111}^{5} + p_{1111}^{6} + 2 p_{1011}^{3} (p_{1111} + 2 p_{1111}^{3}) + p_{1011}^{2} (p_{1111}^{2} + 8 p_{1111}^{4})) \\
  {} = p_{1111} (2 p_{1011}^{2} + 3 p_{1011} p_{1111} + p_{1111}^{2}) (p_{1011}^{3} + p_{1011}^{2} p_{1111} + p_{0110} p_{1111}^{2}).
\end{gather*}
This equation can be resolved for $p_{0110} = f(p_{1011}, p_{1111})$ where
$f$ is a (lengthy) algebraic function involving rational functions of its
arguments and a single square root. The system $\CC T$ together with the
bounds~\eqref{eq:Bounds} define a semialgebraic set and \TT{Mathematica}'s
\TT{FindInstance} function quickly returns a solution typically with large
denominators and an algebraic number of extension degree~2 over~$\BB Q$.
This~distribution proves $\CC L \not\Rightarrow \CIb{XY|ZU} \ge 0$.
A rough map of where such counterexamples lie in the space $\CC T$ is given
in \Cref{fig:Score}.

However, to confirm the Ingleton violation without numerical approximations,
we seek a distribution with \emph{rational} probabilities. The distribution
is rational if $p_{0110}, p_{1011}, p_{1111}$ can be chosen rational, which
hinges on the square root in the algebraic function~$f$ determining $p_{0110}$.
The~term under the square root, expressed in $p_{1011} = \sfrac{a}{b}$ and
$p_{1111} = \sfrac{c}{d}$ with $a,b,c,d \in \BB N$,~reads
\begin{gather*}
  \frac1{b^{8} d^{12}} \Big(
  b^{8} c^{12} + 10 a b^{7} c^{11} d - 2 b^{8} c^{11} d + 41 a^{2} b^{6} c^{10} d^{2} - 16 a b^{7} c^{10} d^{2} + b^{8} c^{10} d^{2} + 88 a^{3} b^{5} c^{9} d^{3} - 46 a^{2} b^{6} c^{9} d^{3} + {} \\
  6 a b^{7} c^{9} d^{3} +  104 a^{4} b^{4} c^{8} d^{4} - 44 a^{3} b^{5} c^{8} d^{4} + 11 a^{2} b^{6} c^{8} d^{4} + 64 a^{5} b^{3} c^{7} d^{5} + 44 a^{4} b^{4} c^{7} d^{5} + 2 a^{3} b^{5} c^{7} d^{5} - {} \\
  2 a^{2} b^{6} c^{7} d^{5} + 16 a^{6} b^{2} c^{6} d^{6} + 136 a^{5} b^{3} c^{6} d^{6} - 6 a^{4} b^{4} c^{6} d^{6} - 14 a^{3} b^{5} c^{6} d^{6} + 112 a^{6} b^{2} c^{5} d^{7} + 26 a^{5} b^{3} c^{5} d^{7} - {} \\
  42 a^{4} b^{4} c^{5} d^{7} + 32 a^{7} b c^{4} d^{8} + 68 a^{6} b^{2} c^{4} d^{8} - 70 a^{5} b^{3} c^{4} d^{8} + a^{4} b^{4} c^{4} d^{8} + 56 a^{7} b c^{3} d^{9} - 68 a^{6} b^{2} c^{3} d^{9} + {} \\
  4 a^{5} b^{3} c^{3} d^{9} + 16 a^{8} c^{2} d^{10} - 36 a^{7} b c^{2} d^{10} + 6 a^{6} b^{2} c^{2} d^{10} - 8 a^{8} c d^{11} + 4 a^{7} b c d^{11} + a^{8} d^{12} \Big).
\end{gather*}
The denominator is always a square, so it suffices to find, in accordance
with \eqref{eq:Bounds}, four positive integers $b \le 160a \le 3b$ and
$d \le 8c \le 3d$ which make the parenthesized numerator into a square.
An~exhaustive search through small denominators $b, d$ turns up $p_{1011}
= \sfrac{2}{99}$ and $p_{1111} = \sfrac{2}{11}$ satisfying this criterion,
because their value
\[
  937\,129\,691\,803\,487\,846\,400 = 30\,612\,574\,080^2
\]
is a perfect square. The resulting rational value $p_{0110} = f(\sfrac{2}{99},
\sfrac{2}{11}) = \sfrac{10}{693}$ does not satisfy \eqref{eq:Bounds} but it still
yields a positive non-Ingleton score. To see this, consider the score $\rho_2$ of
the distribution with the given parameters, write all fractions with their
common denominator $693$ and assemble all terms under one~$\log \sqrt[693]{\blank}$.
Then from
\[
  \left(\exp \rho_2\right)^{693} = \frac{
    24^{24} \cdot 30^{30} \cdot 141^{141} \cdot 168^{168} \cdot 201^{201} \cdot 228^{228} \cdot 294^{294} \cdot 300^{300} \cdot 693^{693}
  }{
    11^{11} \cdot 154^{154} \cdot 198^{198} \cdot 220^{220} \cdot 252^{252} \cdot 308^{308} \cdot 441^{441} \cdot 495^{495}
  }
\]
the violation of the Ingleton inequality is just a matter of comparing the
integers in the numerator and denominator --- a standard task which every
computer algebra system with exact arithmetic on big integers will perform.
The former is approximately $219.148 \cdot 10^{5190}$ and the latter $1.14751
\cdot 10^{5190}$. Thus, the fraction is greater than one and the non-Ingleton
score is positive. Numerically, the score and hence the negative of the
Ingleton expression $\CIb{XY|ZU}$ is approximately~$0.00757$. The distribution
in its entirety is given in the beginning of this note.

\section{Classification of essentially conditional Ingleton inequalities}
\label{sec:Essential}

\subsection{Essential conditionality}

The second part of our theorem concerns essential conditionality,
a notion introduced in \cite{CondInfo}. Given a conditional information
inequality ${\CC L \Rightarrow \CIb{XY|ZU} \ge 0}$ one may ask if it arises
from a valid unconditional information inequality of the form
\begin{equation}
  \label{eq:uI}
  \tag{$\CIbOp_\lambda$}
  \CIb{XY|ZU} + \sum_{\CI{i,j|K} \in \CC L} \lambda_{\CI{i,j|K}} \CId{i,j|K} \ge 0,
\end{equation}
with Lagrange multipliers $\lambda_{\CI{i,j|K}} \ge 0$. The existence of
multipliers which make \eqref{eq:uI} a valid information inequality constitutes
an ``unconditional'' proof of the conditional inequality $\CC L \Rightarrow
\CIb{XY|ZU} \ge 0$; otherwise this inequality is \emph{essentially conditional}.
The masks \eqref{eq:M:1}--\eqref{eq:M:5} show that the conditional Ingleton
inequalities \eqref{eq:11cI}--\eqref{eq:15cI} are in fact not essentially~conditional.
Among the first examples of essentially conditional inequalities due to
Kaced and Romashchenko \cite{CondInfo} are the conditional Ingleton
inequalities \eqref{eq:21cI}--\eqref{eq:24cI}. Hence, the only remaining case
in the classification of essential conditionality for conditional Ingleton
inequalities is the inequality \eqref{eq:25cI} which was recently discovered
by Studený~\cite{StudenyIngleton}.

\begin{remark} \label{rem:Almost}
All unconditional information inequalities are valid for almost-entropic
polymatroids, i.e., points of the closure $\ol{\BO H_n^*}$. This is not
clear for essentially conditional inequalities and \cite[Section~V]{CondInfo}
proves that \eqref{eq:21cI} does not hold almost-entropically but
\eqref{eq:23cI} and \eqref{eq:24cI} do.
\end{remark}

\subsection{Sampling for a counterexample}

If $\lambda$ is a tuple of Lagrange multipliers that makes \eqref{eq:uI}
true and $\mu \ge \lambda$ componentwise, then $\mu$ also makes \eqref{eq:uI}
true since the $\CIdOp$ functionals are non-negative on the entropy region.
Hence there is no loss of generality in assuming that all multipliers are
equal and arbitrarily large but fixed. To prove essential conditionality
we construct counterexamples to \eqref{eq:uI} depending continuously on
$\lambda \to \infty$, i.e., a curve of counterexamples.
The~curves proving essential conditionalities in \cite{CondInfo} all
follow a simple combinatorial recipe:
\begin{enumerate}[label=\arabic*., leftmargin=3em, rightmargin=3em]
\item Commit to state space sizes for all four random variables;
  usually they are all assumed to be binary. This gives rise to
  16 real parameters $P = \Set{ p_{0000}, \dots, p_{1111} }$.
\item Choose a partition of $P$ into four subsets $A, B, C, D$ and
  assign the probabilities
  \begin{center}
  \setlength\tabcolsep{2em}
  \begin{tabular}{c|c|c|c}
    $p \in A$          &    $p \in B$                 &    $p \in C$                  & $p \in D$ \\[1ex] \hline \rule[-2ex]{0pt}{6.5ex}
    $\dfrac1{|A|+|B|}$ & \; $\dfrac1{|A|+|B|} - \eps$ & \; $\dfrac{|B|}{|C|} \, \eps$ & $0$
  \end{tabular}
  \end{center}
  with a real, positive parameter $\eps \to 0$. To ensure that the
  result is a probability distribution we require $|A\cup B| > 0$
  and $|C| > 0$.
\end{enumerate}
A curve of this type converges to a distribution which is uniform on
its support. It is well-known \cite{Quasiunif} that every invalid
information inequality can be refuted by such a distribution --- however,
this result requires unbounded state spaces. The typical argument in
\cite{CondInfo} expands the terms in \eqref{eq:uI} as power series in
$\eps$ around zero and compares convergence orders to conclude that a
small enough value of $\eps$ leads to a violation of the inequality.

Sampling distributions according to the above algorithm and using criteria
based on the limit behavior of the power series coefficients obtained via
\TT{Mathematica}'s \TT{Series} function eventually turns up the following
sparse proof of essential conditionality for \eqref{eq:25cI}:
\begin{gather*}
  p_{0000} = 0, \quad
  p_{0001} = 0, \quad
  p_{0010} = \sfrac15 - \eps, \quad
  p_{0011} = 0, \\
  p_{0100} = 0, \quad
  p_{0101} = 0, \quad
  p_{0110} = \sfrac15, \quad
  p_{0111} = 0, \\
  p_{1000} = 0, \quad
  p_{1001} = 0, \quad
  p_{1010} = \sfrac15, \quad
  p_{1011} = 0, \\
  p_{1100} = \eps, \quad
  p_{1101} = \sfrac15, \quad
  p_{1110} = 0, \quad
  p_{1111} = \sfrac15.
\end{gather*}
The CI~assumptions of \eqref{eq:25cI} are only satisfied in the limit
$\eps = 0$ since
\begin{align*}
  \CId{X,Z|U} = \CId{Y,Z|U} &= \frac15 \log\left(\frac{27}{(3 - 5\eps)^{3-5\eps} \cdot (1 + 5\eps)^{1+5\eps}}\right) \\
  &= \log(3) \, \eps - \frac{10}{3} \, \eps^2 + \frac{100}{27} \, \eps^3 + \mathcal{O}(\eps^4).
\end{align*}
This makes it possible to violate the Ingleton inequality, and indeed:
\begin{align*}
  \CIb{XY|ZU} &= \log\left(\sqrt[5]{\frac{27}{8000}} \cdot \frac{(\sfrac15-\eps)^{\sfrac15-\eps} \cdot (\sfrac45-\eps)^{\sfrac45-\eps} \cdot (\sfrac25+\eps)^{\sfrac25+\eps} \cdot \eps^\eps}{(\sfrac25-\eps)^{2(\sfrac25-\eps)} \cdot (\sfrac35-\eps)^{\sfrac35-\eps} \cdot (\sfrac15+\eps)^{3(\sfrac15+\eps)}} \right) \\
  &= (\log(30 \eps) - 1) \, \eps - \frac{155}{25} \, \eps^2 + \frac{11525}{864} \, \eps^3 + \mathcal{O}(\eps^4).
\end{align*}
The expression \eqref{eq:uI} in our case is
\begin{align*}
  \CIb{XY|ZU} + \lambda (\CId{X,Z|U} + \CId{Y,Z|U}) = (-1 + 2 \lambda \log(3) + \log(30 \eps)) \, \eps + \mathcal{O}(\eps^2)
\end{align*}
whose $\eps$-order coefficient tends to $-\infty$ as $\eps \to 0$ for any fixed
$\lambda$. Hence, every unconditional version of \eqref{eq:25cI} can be violated
on our curve of distributions, which proves essential conditionality.

\section{Remarks}
\label{sec:Remarks}

\begin{paraenum}

\item
The distribution constructed in \Cref{sec:Construction} satisfies the four
CI~statements in $\CC L$ and none other. This~can be checked computationally
but it also follows from \Cref{sec:CondIng} since every superset of $\CC L$
implies the Ingleton inequality.

\item
The entropy vector of that distribution is a conic combination of twelve
extreme rays of~$\BO H_4$ (corresponding to the twelve coatoms in the
lattice of semimatroids above $\CC L$; cf.~\cite{MatusI}). The~only ray
which violates the Ingleton inequality is not entropic. Thus, our
construction gives an \emph{entropic} conic combination of these not
necessarily entropic polymatroids where the non-Ingleton component has
sufficiently high weight.

\item
All counterexamples to potential conditional Ingleton inequalities with
inclusion-minimal assumptions \cite[Section~IV.B]{StudenyIngleton} as well
as all proofs of essential conditionality \cite[Section~IV.A]{CondInfo}
require only rational binary distributions. This is remarkable insofar
as there exist CI~inference rules which are valid for binary random vectors
but not in general; see \cite{MatusForks}. Whether every wrong CI~inference
rule can be refuted by a rational distribution is equivalent to
\cite[Conjecture]{MatusIII} and still open.

\item
The method of \cite{MatusForks} to construct binary distributions with
prescribed CI~structure using the Fourier--Stieltjes transform even produces
distributions close to the uniform distribution. This allows one to concentrate
on satisfying the CI~equations only, because every binary tensor close to the
uniform distribution has strictly positive entries and thus yields a positive
probability distribution after multiplying all entries by a normalizing
constant.
Matúš's parametrization of the model $\CC L_2$ depends on a solution to the
associated \emph{solvability system} whose components appear as exponents of
the parameters. The smallest integral solution to the solvability system is
$(x_{12}, x_{13}, x_{14}, x_{23}, x_{24}, x_{34}) = (1,2,1,1,2,1)$; see
\cite[Theorem~1]{MatusForks} for details. In the nomenclature of this theorem
(and its proof), the non-Ingleton score is then given~by
\[
  (\gamma^2+1) \log(\gamma^2+1) + \sfrac12 (\gamma-1) \log(\gamma-1)
  - (\gamma^2-1) \log(\gamma^2-1) - \sfrac12 (\gamma+1) \log(\gamma+1)
\]
for $\gamma$ small but positive.
This function in $\gamma$ has one root in the interval $(0,1)$ where
it passes from negative on the left to positive values on the right.
The root has the approximate value of $0.72766$.
Using cylindrical algebraic decomposition in \TT{Mathematica},
it can be verified that Matúš's construction
does not produce tensors with non-negative entries (ergo probability
distributions) if $\gamma > 0.727$ is imposed.
It remains open whether there exist counterexamples to the validity
of the Ingleton inequality subject to $\CC L$ and arbitrarily close
to uniform or even just without zero entries.

\item The same method applies to the search for a proof of essential
conditionality in \Cref{sec:Essential} because the CI~assumptions
$\CI{X,Z|U} \wedge \CI{Y,Z|U}$ have conditioning sets of size~one.
Moreover, this statistical model has a rational parametrization:
its conditionals with respect to $U$ belong to the \emph{marginal
independence model} $\CI{X,Z} \wedge \CI{Y,Z}$ which has a monomial
parametrization in Möbius coordinates by \cite{MarginalIndependence}.
Lastly, the entropy vectors arising from those distributions in the
marginal independence model which have no private information have
been completely characterized by \cite{MatusPiecewise}. The random
search carried out in \Cref{sec:Essential} found a counter\-example
more quickly than any of these approaches.

\item
Combinatorial and group-theoretic constructions of distributions with
large violations of the Ingleton inequality have been investigated in
\cite{LargeViolations} in the context of the four-atom conjecture,
which was then refuted in \cite{MatusCsirmaz}.

\item
The last part of Open Question 2 in \cite{StudenyIngleton} concerns
validity of \eqref{eq:21cI}--\eqref{eq:25cI} for almost-entropic points.
As mentioned in \Cref{rem:Almost} some cases are settled in \cite{CondInfo}
with different answers. The status of \eqref{eq:22cI} and of \eqref{eq:25cI}
is open.

\end{paraenum}

\subsubsection*{Acknowledgements}
I thank the anonymous referee for hints for improving the presentation.
I~would also like to thank Mima Stanojkovski and Rosa Winter
for their immediate interest, code samples and an inspiring discussion
about finding rational points on varieties --- even though the brute force
approach turned out to succeed more quickly this time.

\pagebreak

\bibliographystyle{tboege}
\bibliography{milan1.bib}
\nocitesoft{M2,4ti2,Normaliz,CaDiCaL,Mathematica}
\bibliographystylesoft{tboege}
\bibliographysoft{milan1.bib}
\enlargethispage{4em}

\end{document}